\documentclass[aps,preprint,superscriptaddress,showpacs]{revtex4}

\usepackage{graphicx,epsfig}

\begin{document}

\title{Generating Black Strings in Higher Dimensions}
\author{L. A. L\'opez$^1$, A. Feinstein$^2$ and N. Bret\'on$^{}$}
\affiliation{$^{}$ Dpto. de F\'{\i}sica, Centro de Investigaci\'on 
y de Estudios Avanzados del I. P. N.,
Apdo. 14-740, D.F., M\'exico.\\
$^2$ Dpto. de F\'{\i}sica Te\'orica, Universidad del Pa\'{\i}s
Vasco, Apdo. 644, E-48080, Bilbao, Spain.}

\begin{abstract}
Starting with a Zipoy-Voorhees line element we construct and study the
three parameter family of solutions describing a deformed black string
with arbitrary tension.
\end{abstract}

\pacs{04.50.Gh, 11.10.Kk, 11.25.-w}

\maketitle

\section{Introduction}
   
Associated with string theory and higher dimensional universes, there has
been a renewed interest in higher dimensional solutions to Einstein field
equations, among the most interesting being black rings \cite{b01}, KK
bubbles as well as black strings.

One simple way to lift the 4D Schwarzschild black hole to a 5D black
string is to add an extra flat dimension. In other words it is possible to
uniformly extend the 4D black hole with $S^{2}$ horizon into the fifth
dimension producing a hypercylindrical black hole $S^{2} \times R$
\cite{GL}. A striking fact related to black strings is their generic
instability and final fate and whether these end up as black holes or
different objects is still not well understood \cite{GL2}, \cite{Harmark2}. 
In this sense
it is interesting testing more general black string solutions both static
and non-static. Some generalizations are obtained by applying a boost to
the string solution, dotating the string with the momentum along the fifth
coordinate \cite{Kim}. In this work we derive and report a generalized
black string solution with mass, arbitrary tension and an additional
parameter related to deformation from spherical symmetry.
  
To generate a generalized black string solution we take as seed the
Zipoy-Vorhees (ZV) family of solutions and then lift it to 5D by the
procedure introduced in \cite{FVM},\cite{b5}. Particular cases are the
static black string with arbitrary tension of C. H. Lee \cite{CHLee} and
the Gregory-Laflamme black string \cite{GL}. Trapped surfaces and
horizons, as well as the mass and string tension parameters, are analyzed
and discussed.

\section{Generalized Black String}

We briefly explain the idea of the generating technique
(see details in \cite{b5}). Let us consider the 5D vacuum Einstein action,

\begin{equation}
S=- \frac{1}{16 \pi {\hat G}} \int{ {\hat R} \sqrt{-{\hat g}} d^4x dy }
\label{S5D}
\end{equation}

where the hatted quantities mean the fifth dimensional version of
the scalar curvature $R$, the metric tensor $g$ and the gravitational Newton 
constant $G$ and $y$ is the fifth coordinate. 
The equations of motion derived of the variation of action (\ref{S5D}) coincide 
with the Einstein-scalar coupled equations in four dimensions.
For a static line element given in Weyl coordinates, 

\begin{equation}
ds^{2}=-e^{2U}dt^{2}+e^{-2U+2\sigma}(d\rho^{2}+dz^{2})+
\rho^{2}e^{-2U}d\phi^{2},
\label{A1}
\end{equation}

and a minimaly coupled scalar field $\varphi$, these field equations are given by

\begin{eqnarray}
0&=& U_{,\rho\rho}+\frac{1}{\rho}U_{,\rho}+U_{,zz}, \nonumber\\
0&=& \varphi_{,\rho\rho}+\frac{1}{\rho}\varphi_{,\rho}+\varphi_{,zz}, \nonumber\\ 
\sigma_{, \rho}&=& \rho ( U_{,\rho}^2- U_{,z}^2) + \frac{\rho}{2} (\varphi_{,\rho}^2 -\varphi_{,z}^2), \nonumber\\
\sigma_{,z}&=& 2 \rho  U_{,\rho}U_{,z} + 2\rho \varphi_{,\rho}\varphi_{,z}.
\label{field_eqs}
\end{eqnarray}

Let us consider the ZV solution \cite {b3} in Weyl coordinates (for a
discussion on Zipoy-Voorhees solution see \cite {b4}), with the line element
(\ref{A1}) and the metric functions given by:

\begin{eqnarray}
U&=&\frac{1}{2}\ln\left[\frac{r_{+}+r_{-}-2m}{r_{+}+r_{-}+2m}\right]^
{\frac{q}{2}}, \nonumber\\
\sigma&=&\frac{1}{2}\ln\left[\frac{(r_{+}+r_{-})^{2}-4m^{2}}{4r_{+}r_{-}}
\right]^{\frac{q^{2}}{4}}, \nonumber\\ r_{\pm} ^{2}&=&\rho^{2}+(z \pm
m)^{2}.
\label{A1.1}
\end{eqnarray}
  
ZV solution is characterized by two free parameters $m$ and $q$ related to
mass and spherical-symmetry deformation, respectively.  For $q=2$ we
recover Schwarzschild solution while if $q=4$ it represents the Darmois
solution \cite{b3}.

Adding the scalar field in order to use the generating algorithm of
\cite{FVM},\cite{b5}

\begin{equation}\label{A2}
\varphi=\frac{A}{2}\ln\left[\frac{r_{+}+r_{-}-2m}{r_{+}+r_{-}+2m}\right],
\end{equation} 
where the scalar field is a solution of Laplace's equation
$\varphi_{,\rho\rho}+\frac{1}{\rho}\varphi_{,\rho}+\varphi_{,zz} =
0$, we obtain the spacetime with a scalar field $\varphi$,
expressed as

\begin{equation}\label{A3}
ds^{2} _{sc}=-e^{2U}dt^{2}+e^{-2U+2(\sigma +
\sigma_{sc})}(d\rho^{2}+dz^{2})+\rho^{2}e^{-2U}d\phi^{2}
\end{equation} with
 
\begin{equation}\label{A4}
\sigma_{sc}=\frac{A^{2}}{2}\ln\left[\frac{(r_{+}+r_{-})^{2}-4m^{2}}
{4r_{+}r_{-}}\right].
\end{equation}

The above solution is an exact solution of the Einstein-scalar
equations (\ref{field_eqs}).

Using now the method described in \cite{b5}, and considering the metric
(\ref{A3}) as seed, we lift the solution to five dimensions:
 
\begin{equation}\label{A5}
dS_{5} ^{2}=
e^{-\frac{2\varphi}{\sqrt{3}}}(ds_{sc}^{2})+e^{^{\frac{4\varphi}{\sqrt{3}}}}
d\omega^{2},
\end{equation} 
explicitly

\begin{equation}\label{A6}
dS^{2}_{5}=-H^{\frac{q}{2}-\frac{A}{\sqrt{3}}}dt^{2}
+H^{-\frac{q}{2}-\frac{A}{\sqrt{3}}}\rho^{2}d\phi^{2}
+ H^{ \frac{2A}{\sqrt{3}}}d\omega^{2}
+ H^{-\frac{q}{2}-\frac{A}{\sqrt{3}}}
F^{\frac{q^{2}}{4} + A^{2}}(d\rho^{2} + dz^{2}),
\end{equation} where 

\begin{equation}\label{A6.1}
H=\frac{(r_{+}+r_{-}-2m)}{(r_{+}+r_{-}+2m)}, \quad 
F=\frac{(r_{+}+r_{-})^{2}-4m^{2}}{4r_{+}r_{-}}.
\end{equation}

Transforming the metric (\ref{A6}) to the Schwarzschild-like
coordinates $(r, \theta)$, with

\begin{equation}\label{A7}
\rho^{2}=\left(1-\frac{2m}{r}\right)r^{2} \sin ^{2} \theta, \;\;\;\;\
z=r\left(1-\frac{m}{r}\right)\cos \theta ,
\end{equation} we obtain:

\begin{eqnarray}\label{A8}
dS^{2}_{5}&=&-G^{\frac{q}{2}-\frac{A}{\sqrt{3}}}dt^{2}
+G^{1-\frac{q}{2}-\frac{A}{\sqrt{3}}}
r^{2}\sin^{2} \theta d \phi^{2} + G^{\frac{2A}{\sqrt{3}}}
d\omega^{2} \nonumber\\ 
&& + \left( {G + \frac{m^{2}\sin^{2}\theta}{r^2}} \right)^{1-A^{2}-
\frac{q^{2}}{4}}
G^{P}r^2
\left( G^{-1} \frac{dr^{2}}{r^{2}}+d\theta^{2}\right),
\end{eqnarray}
where $G=(1-2m/r)$ and
$P=A^{2}+\frac{q^{2}}{4}-\frac{A}{\sqrt{3}}-\frac{q}{2}$.
  
This solution is a generalized 5D black string characterized by three free
parameters: $A$, $q$ and $m$.  Note that to recover transverse spherical
symmetry, we must choose $A^{2}+\frac{q^{2}}{4}=1$ in (\ref{A8}). Doing so
the term with $\sin^{2} \theta$ does not appear in $g_{rr}$ and
$g_{\theta\theta}$, obtaining an element proportional to $r^{2}d\Omega$:

\begin{equation}\label{sph}
dS^{2}_{5}=-G^{\frac{q}{2}-\frac{A}{\sqrt{3}}}dt^{2}+G^{1-\frac{q}{2}-
\frac{A}{\sqrt{3}}}
r^{2}(d \theta^2 + \sin^{2} \theta d \phi^{2}) + G^{\frac{2A}{\sqrt{3}}}
d\omega^{2} + G^{-\frac{q}{2}-\frac{A}{\sqrt{3}}}dr^{2}.
\end{equation}

The case $A=0$ and $q=2$ corresponds to the black string \cite{GL},

\begin{equation}\label{A19}
dS_{5}^{2}=-\left(1-\frac{2m}{r}\right)dt^{2}+
\left(1-\frac{2m}{r}\right)^{-1}dr^{2}+r^{2}(d\theta^{2}+\sin^{2}\theta
d\phi^{2})+d\omega^{2}.
\end{equation}

In the case
$A^{2}=3/4$ and $q=1$ solution (\ref{A8})
becomes

\begin{equation}\label{KK}
dS^{2}_{}=
[r^2 (d\theta^2+ \sin^2{\theta} d \phi^2) + G^{-1} dr^2+ G d \omega^2] -dt^{2},
\end{equation}
 
having in square brackets the Schwarzschild line element with the
euclidean signature . By analytic continuation of the 2-sphere in
(\ref{KK})  to a 2-de Sitter making $(\theta - \pi/2) \mapsto i \tau$ and
$t \mapsto i \psi$ we get the expression of the 5D KK-bubble of nothing
\cite{bubnot}

\begin{equation}
ds^2=[r^2 ( -d\tau^2+ \cosh^2{\tau} d \phi^2) + G^{-1} dr^2+ G d \omega^2] +
d \psi^{2}.
\end{equation} 

Another interesting particular case is the black string with arbitrary
tension \cite{CHLee}. Performing in (\ref{sph}) the coordinate
transformation

\begin{equation}\label{A21}
r=\varrho \left(1+\frac{k}{\varrho}\right)^{2} ,\qquad
{\rm with} \quad m=2k, 
\end{equation} 
the metric is transformed into:

\begin{equation}\label{A22}
dS^{2}_{5}=D^{-\frac{2}{\sqrt{3}}A}[-D^{q}dt^{2}+(1-k^2/\varrho^2)^2D^{-q}
(d\varrho^{2}+\varrho^{2}d\theta^{2}+
\varrho^{2} \sin^{2}\theta d\phi^{2})] 
+ D^{\frac{4}{\sqrt{3}}A}d \omega^{2},
\end{equation} 
where $D=(1-k/\varrho)/(1+k/\varrho)$.
Parametrizing with $(s,a)$ related to $(A,q)$ by

\begin{equation}\label{A24}
2A=\frac{2a-1}{\sqrt{1-a+a^{2}}},  \quad
q=\frac{3}{\sqrt{3(1-a+a^{2})}}, \quad -\frac{2}{\sqrt{3}}A+q=s,
\quad s=\frac{2(2-a)}{\sqrt{3(1-a+a^{2})}},
\end{equation}
the metric (\ref{A22}) is written as

\begin{equation}\label{A23}
dS^{2}_{5}=-D^{s}dt^{2}+
(1-k^2/\varrho^2)^2
D^{-\frac{1+a}{2-a}s}
(d\varrho^{2}+\varrho^{2}d\theta^{2}+ \varrho^{2}sen^{2}\theta
d\phi^{2})+ D^{\frac{1-2a}{2-a}s} d \omega^{2}.
\end{equation}
  
In this parametrization the solution has only two free parameters $(k, a)$
and we identify it as the black string with arbitrary tension previously
derived in \cite{CHLee} and studied in \cite{CHLee2}. Notice that we have
given a much simpler derivation of the solution.

\section{Properties of the 5D generalized black string}

\begin{figure}
\centering
\includegraphics[width=8.6cm,height=4cm]{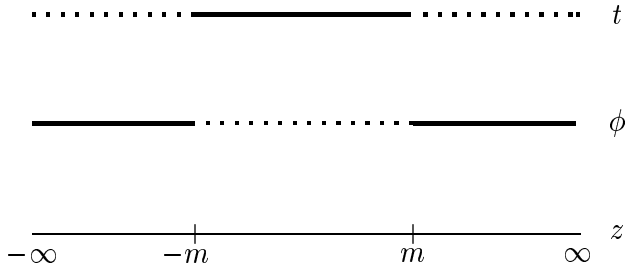}
\caption{\label{fig1}
Rod structure of the 4D Zipoy-Vorhees solution}
\end{figure}


\begin{figure}
\centering
\includegraphics[width=8.6cm,height=4cm]{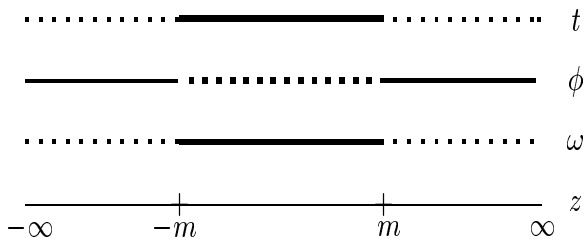}
\caption{\label{fig2}
Rod structure of the 5D black string generated from Zipoy-Vorhees and
inserting a finite rod with scalar charge $2A / \sqrt{3}$ in the interval
$(-m, m)$ in the fifth dimension. The bold dotted line along
$\partial_{\phi}$ is a rod with negative mass density (conical
singularity).}
\end{figure}
 

\subsection{Rod-structure}
   
It is illustrative to sketch briefly how the generation method works in
terms of the rod-structure \cite{Harmark}, \cite{Harmark3}.  
The ZV seed solution, Eqs.
(\ref{A1})-(\ref{A1.1}), is characterized by the rod-structure shown in
Fig.  \ref{fig1}.  It differs from Schwarzschild only in the mass density
along the Killing directions: for Schwarzschild the density in the
interval $[-m,m]$ along the $\partial_{t}$ direction is 1, while for ZV
the density is $q/2$. In the intervals $[- \infty, -m]$ and $[m,\infty]$
along $\partial_{\phi}$ the density is also $q/2$ ($q=2$ for
Schwarzschild).
 
Introducing the scalar field $\varphi$ and then lifting to 5D in terms of
the rod structure amounts to introduce into the fifth dimension a rod of
density $2A/ \sqrt{3}$.  The rod structure of the 5D black string, Eqs.
(\ref{A6})-(\ref{A6.1}), is (shown in Fig.  \ref{fig2} ) the following:
   
There is a density $2A/ \sqrt{3}$ in the interval $[-m,m]$ along the
$\partial_{w}$ direction. A conical singularity along the
$\partial_{\phi}$ direction is shown as the negative density $-A/
\sqrt{3}$ in the interval $[-m,m]$ (bold dotted line); also there is a
positive density of $q/2$ in the intervals $[-\infty,m]$ and $[m,\infty]$
along the $\partial_{\phi}$ direction. The density is $(q/2-A/ \sqrt{3})$
in the interval $[-m,m]$ along the $\partial_{t}$ direction; to have a
physical horizon, if one wants to keep calling the system a black string,
the density along the timelike direction should be positive, i.e. 
$(q/2-A/ \sqrt{3})>0$.


\subsection{Apparent horizons and singularities}

The invariant that defines trapped or marginally trapped surfaces for the
generated solution (\ref{A8}) is analyzed in what follows.

Trapped surfaces $S_{X^{a}}$ and located horizons corresponding to the
spacetime (\ref{A8}), are determined by the scalar $\kappa$ ( see
\cite{b6}). To calculate it, we fix the coordinates $x^{a}=\{r,t\}$ and
denote the local coordinates on the surface $S_{X^{a}}$ by
$x^{A}=\{\theta,\phi,\omega\}$. The function $e^{V} \equiv \sqrt{\det
g_{AB}}, (A,B=\theta,\phi,\omega)$, gives the canonical volume element of
the surfaces $S_{X^{a}}$. Introducing $\textbf{g}_{a}\equiv g_{aA}dx^{a},
(a=r,t)$, $H_{\mu}= \delta_{\mu} ^{a} (V_{,a}-div \overrightarrow{g_{a}})$
and defining

\begin{equation}\label{A9}
\kappa_{x^{a}}= -g^{ab}H_{b}H_{c}|_{S_{X^{a}}},
\end{equation} the invariant $\kappa$ 
for the obtained solution amounts to

\begin{equation}\label{A10}
\kappa_{ \{r,t \}} = -g^{rr}V_{,r}^{2}.
\end{equation}

For the solution of our interest,
\begin{equation}\label{A11}
V=\frac{1}{2} \ln
\left[\frac{r^{2(A^{2}+\frac{q^{2}}{4}+1)}\sin^{2}\theta 
(1-\frac{2m}{r})^{1-q+A^{2}+\frac{q^{2}}{4}}}{(r^{2}-2mr +  
m^{2}\sin^{2}\theta)^{A^{2}+\frac{q^{2}}{4}-1}}\right]
\end{equation} and therefore, $\kappa$ is obtained as

\begin{equation}\label{A12}
\kappa=-\frac{\{[r-m(1+\cos\theta)][r-m(1-\cos\theta)]\}^{A^{2}+
\frac{q^{2}}{4}-3}[{\tilde V}_{,r}]^{2}}
{r^{A^{2}+\frac{q^{2}}{4}+\frac{A}{\sqrt{3}}+\frac{q}{2}+1}
(r-2m)^{A^{2}+\frac{q^{2}}{4}-\frac{A}{\sqrt{3}}-\frac{q}{2}+1}}
\end{equation} with

\begin{eqnarray}\label{A13}
{\tilde V}_{,r}
&=& 2r^3-(6+q)mr^2+[4+2q+(1+A^2+ \frac{q^{2}}{4}) \sin^2{\theta}]m^2r-
(1+A^2+ \frac{q^{2}}{4}+q) \sin^2{\theta}m^3 \nonumber\\
&=&(r-m)[2r(r-2m)+m^{2}\sin^{2}\theta(A^{2}+
\frac{q^{2}}{4}+1)]-mq[r^{2}-2mr+m^{2}\sin^{2}\theta],
\end{eqnarray}

Let us analyze the marginally trapped surfaces defined when $\kappa=0$ as
well as the singularities if $\kappa$ diverges.  To determine those values
of $\kappa$ cancelations in the denominator coming from ${\tilde V}_{,r}$
should be taken into account.

Assuming that $A>0$ and $q>0$, then there exists a singularity at $r=0$
since the exponent of $r$ in the denominator of (\ref{A12}) is always
positive, $A^{2}+\frac{q^{2}}{4}+\frac{A}{\sqrt{3}}+\frac{q}{2}+1 >0$.

Also there are two marginally trapped surfaces at $r_{\pm} =m(1 \pm
\cos\theta)$, nevertheless these are wrapped or hidden by the naked
singularity located at $r=2m$. Depending on the values of $A$ and $q$, in
the product of ${\tilde V}_{r}$ and $g^{rr}$ may be cancelations that
define horizons that coincide with the naked singularity $r=2m$. Another
kind of marginally trapped surfaces that overlap the singularity in some
regions, arise from the real root $r_h(m,A,q,\theta)$ of the third degree
polynomial ${\tilde V}_{r}$. These surfaces when $r_h>2m$ do not cover
completely the singularity due to the dependence on $\theta$. Then the
situation when $r_h>2m$ is that for some ranges of $\theta$ the
singularity is hidden but coming from other direction the singularity is
naked, i.e. the singularity is partially naked (or partially dressed).

For the black string with arbitrary tension ($A^2+q^2/4=1$), $\kappa$
amounts to

\begin{equation}\label{A25}
\kappa=-\frac{4}{r^{2}}\left(1-\frac{2m}{r}\right)^
{\frac{A}{\sqrt{3}}+\frac{q}{2}-2}\left(1-\frac{m(1+q/2)}{r}\right)^{2},
\end{equation}
 
The solution is singular at $r=2m$ unless $q=2$ ($a=1/2$), case in which
it possesses a horizon at $r=2m$ ($r=4k$). In any other case ($q<2$) there
is a naked singularity at $r=2m$ that hides horizons at $r=m(1+q/2)$. The
$\kappa$ diverges at $r=0$ showing the singularity there.

In the case $A=0$ and $q=2$ that we recover the black string solution
\cite{GL}, it possesses a horizon at $r=2m$ as the expression for $\kappa$
shows,

\begin{equation}\label{A20}
\kappa=-\left(1-\frac{2m}{r}\right)\left(\frac{2}{r}\right)^{2}
\end{equation}

In this case the naked singularity disappears and instead a horizon is
located at $r=2m$.

With respect to the analysis of the singularities we  have attempted to draw
Penrose diagram to dilucidate the global structure of the solutions.
However, due to the reduced symmetry of  the line element (\ref{A8})  the
metric function $g_{rr}$ does not depend
only on $r$ but also on $\theta$. Even if one drops the $\theta$ dependence,
by fixing the angle, the
integration to be performed involve hypergeometric functions, and the
analysis by this
method does not produce a clear picture. On the other hand, the information
one obtains by
studying the scalar $\kappa$, is essentially the same that can be obtained
from the Kretschmann scalar.

\subsection{Mass and string tension}

To characterize the 5D black string we need to determine the mass and
tension of the source using the asymptotic expansion of the metric
functions $g_{tt}$ and $g_{\omega \omega}$ of (\ref{A8}),

\begin{eqnarray}\label{A14}
-g_{tt}&=&\left(1-\frac{2m}{r}\right)^{-\frac{A}{\sqrt{3}}+\frac{q}{2}} 
\nonumber\\
&& \simeq
1+\left(\frac{A}{\sqrt{3}}-\frac{q}{2}\right)\frac{2m}{r}+\frac{1}{2} 
\left(\frac{A}{\sqrt{3}}-\frac{q}{2}\right)\left(\frac{A}{\sqrt{3}}
-\frac{q}{2}+1\right)\frac{4m^{2}}{r^{2}}+ \cdots \\
g_{\omega\omega}&=&\left(1-\frac{2m}{r}\right)^{\frac{2A}{\sqrt{3}}}  
\simeq 1-
\frac{2A}{\sqrt{3}}\frac{2m}{r}+\frac{2A}{\sqrt{3}}\left(\frac{2A}{\sqrt{3}}
-1\right)\left(\frac{4m^{2}}{r^2}\right)+
\cdots
\end{eqnarray}

Comparing with the asymptotic form of metric around a stationary
matter source as given in \cite{Sorkin}, \cite{Harmark4}:

\begin{eqnarray}\label{A16}
g_{tt} && \simeq -1+\frac{4G_{5}M(2-a)}{3r}, \\
g_{\omega\omega} && \simeq 1+\frac{4G_{5}M(1-2a)}{3r},
\end{eqnarray}

The mass and string tension of the 5D black string (\ref{A8})
are identified then as (we are considering $G_{5}=1$)

\begin{equation}\label{A18}
M=\frac{2qm}{4G_{5}}, \quad
Ma=\tau=\frac{m}{4G_{5}}\left(\frac{6A}{\sqrt{3}}+q \right).
\end{equation}
  
Therefore when (\ref{A1}) is lifted to 5D, by means of the scalar field
$\varphi (A,r,m)$, the string tension changes in $\frac{6A}{\sqrt{3}}$,
that can modify the tension and even make it vanish by calibrating $A=-
\sqrt{3}q/6$, however, from the rod-structure we learned that $A>q/2$ in
order to have a horizon.

In the case $A=0$ and $q=2$ we recover the black string solution
\cite{GL}, with $\tau=M/2$. The case $A= \sqrt{3}/2$ and $q=1$ corresponds
to a black string with $\tau=2M$.


\section{Summary}

In this paper by equiping first the 4D Zipoy-Vorhees solution with a
scalar field, we lifted it to five dimensions; the so generated 5D
solution is a black string characterized by three free parameters: mass,
tension and deformation parameter.  There is a physical spacetime
singularity at $r=0$. Moreover, the horizons do not cover completely the
naked singularity at $r=2m$, except in the case that $r=2m$ is the horizon
itself. Therefore, in 5D the cosmic censorship conjecture does not hold.
 
The static black string with arbitrary tension is a particular case when
there is transversal spherical symmetry.  The stability of the 5D
generalized black string deserves further analysis, particularly to
acquaint how the deformation parameter $A$ can affect the stability of the
solution, since it modifies the string tension. 
We have seen that the three parameter generalizations of the black string
solutions are singular. This may well be an indicator that the black strings
are unstable in the sense of \cite{GL2}, \cite{Harmark2}.

Interesting solutions are
also obtained by an analytic continuations of the ones obtained in this
report. These and other issues will be discussed elsewhere.

\begin{acknowledgments}
A.F.  acknowledges the support of the Basque Government Grant
GICO7/51-IT-221-07 and The Spanish Science Ministry Grant FIS2007-61800.
N. B. would like to thank the colleagues of UPV/EHU for warm hospitality.
L. A. L\'opez acknowledges Conacyt-M\'exico for a Ph. D. grant. Partial
support of Conacyt-Mexico Project 49182-F is also acknowledged.
\end{acknowledgments}


\begin{thebibliography}{99}
\bibitem{b01} R. Emparan and H. S. Reall, \textit{
Black holes in higher dimensions}, 
review for Living Reviews in Relativity, 
arXiv:0801.3471

\bibitem{GL} R. Gregory and R. Laflamme, \textit{Hypercylindrical black
holes}, Phys. Rev. D {\bf37}, 305 (1988)

\bibitem{GL2} R. Gregory and R. Laflamme, \textit{Black strings and
p-branes are unstable}, Phys. Rev. Lett. {\bf 70}, 2837 (1993).

\bibitem{Harmark2}
T. Harmark, V. Niarchos and N. A. Obers, 
\textit{Instabilities of black strings and branes},  
Class. Quantum Grav. {\bf24}, R1-R90 (2007)

\bibitem{Kim}
H. C. Kim and J. Lee, \textit{Extraordinary vacuum black string solution,}
Phys. Rev. D {\bf 77}, 024012 (2008)

\bibitem{FVM} A. Feinstein, M. A. Vazquez-Mozo, \textit{M theory
resolution of four-dimensional cosmological singularities}, Nucl. Phys
{\bf B 568}, 405, (2000), hep-th/9906006

\bibitem{b5} N. Bret\'on, A. Feinstein and L.A. L\'opez,
\textit{Generating generalized $G_{D-2}$ solutions}, Phys. Rev. D {\bf 77},
124021 (2008)

\bibitem{CHLee} C. H. Lee, \textit{Black String Solution with Arbitrary
Tension}, Phys. Rev. D {\bf74} 104016 (2006)

\bibitem{b3} H. Stephani, D. Kramer, M. MacCallum, C. Hoenselaers and E. Herlt,
\textit {Exact Solutions of Einstein's Field Equations}, Cambridge University
 Press, Second Ed., (2003).

\bibitem{b4} L. Fern\'andez-Jambrina, \textit{ Moment density of Zipoy's
dipole solution}, Class. Quantum Grav. {\bf11}, 1483 (1994)

\bibitem{bubnot}
O. Aharony, M. Fabinger, G. T. Horowitz and E. Silverstein,
{\it Clean Time-dependent String Backgrounds from Bubble Baths},
JHEP, 0207 (2002)007, arXiv: hep-th/0204158.

\bibitem{CHLee2}
I. Cho, G. W. Kang, S. P. Kim and  C. H. Lee, \textit{
Spacetime structure of 5D hypercylindrical vacuum solutions with tension},
J. Korean Phys. Soc. {\bf 53} 1089 (2008). arXiv:0709.1021

\bibitem{Harmark}
T.~Harmark, \textit{Stationary and axisymmetric solutions of
higher-dimensional general relativity}, Phys.\ Rev.\ D {\bf 70}, 124002
(2004). 

\bibitem{Harmark3}
T.~Harmark and P. Olesen, \textit{Structure of stationary and
axisymmetric metrics}, Phys.\ Rev.\ D {\bf 72}, 124017 (2005).

\bibitem{b6} J. M. M. Senovilla, \textit{ Trapped surfaces, horizons and
exact solution in higher dimensions}, Class. Quantum Grav. {\bf19}, L113
(2002)

\bibitem{Sorkin} B. Kol, E. Sorkin and T. Piran, \textit{Caged black holes:
Black holes in compactified spacetimes. I. Theory}
Phys. Rev. D {\bf 69}, 064031 (2004).

\bibitem{Harmark4}
T. Harmark and N. A. Obers, 
\textit{New phase diagram for black holes and strings on cylinders},  
Class. Quantum Grav. {\bf 21}, 1709 (2004)


\end{thebibliography}
\end{document}